\documentclass[aps,prl,reprint,a4paper,superscriptaddress,amsmath,amsfonts,amssymb, longbibliography]{revtex4-1}
\usepackage{bm}
\usepackage[caption=false]{subfig}
\usepackage{tikz}
\usetikzlibrary{shapes.misc,positioning,shapes}
\usetikzlibrary{patterns}
\usepackage{array, multirow}
\usepackage[colorlinks,linkcolor=blue,citecolor=blue]{hyperref}%
\usepackage{relsize}

\usepackage{times}
\usepackage{xcolor,colortbl}

\newcolumntype{P}[1]{>{\centering\arraybackslash}p{#1}}

\begin{document}
\title{Measurement Based Feedback Quantum Control With Deep Reinforcement Learning for a Double-well Non-linear Potential}
\date{\today}

\author{Sangkha Borah}
\email{sangkha.borah@oist.jp}
\affiliation{Okinawa Institute of Science and Technology Graduate University, Onna-son, Okinawa 904-0495, Japan}
\author{Bijita Sarma}
\affiliation{Okinawa Institute of Science and Technology Graduate University, Onna-son, Okinawa 904-0495, Japan}
\author{Michael Kewming}
\affiliation{Centre for Engineered Quantum Systems, School of Mathematics and Physics, University of Queensland, QLD 4072 Australia}
\author{Gerard J. Milburn}
\affiliation{Centre for Engineered Quantum Systems, School of Mathematics and Physics, University of Queensland, QLD 4072 Australia}
\author{Jason Twamley}
\affiliation{Okinawa Institute of Science and Technology Graduate University, Onna-son, Okinawa 904-0495, Japan}

\begin{abstract}
Closed loop quantum control uses measurement to control the dynamics of a quantum system to achieve  either a desired target state or target dynamics. In the case when the quantum Hamiltonian is quadratic in ${x}$ and ${p}$,  there are known optimal control techniques to drive the dynamics towards particular states e.g. the ground state. However, for nonlinear Hamiltonians such control techniques often fail. We apply  Deep Reinforcement Learning (DRL), where an artificial neural agent explores and learns to control the quantum evolution of a highly non-linear system (double well), driving the system towards the ground state with high fidelity. We consider a DRL strategy which is particularly motivated by experiment where the quantum system is continuously but weakly measured. This measurement is then fed back to the neural agent and used for training. We show that the DRL can effectively learn counter-intuitive strategies to cool the system to a nearly-pure `cat' state which has a high overlap fidelity with the true ground state. 
\end{abstract}
\maketitle
As the research on quantum communication and computation has progressed rapidly with the goal of achieving the holy grail of quantum computing, quantum state engineering has begun to take on a high profile~\cite{Cirac2009Sep,Motta2020, Love2020}. Of particular importance are feedback control techniques, in which a physical system subjected to noise is continuously monitored in real time while using measurement information to impart specific driving controls to modulate the system dynamics~\cite{wiseman_milburn_book}. Unlike classical systems, measurement control of a quantum mechanical system is challenging for a number of reasons. Firstly, the act of continuously observing a quantum system introduces non-linearity within the conditioned dynamics. Secondly, continuous measurement on a quantum system generally
alters it, generating measurement-induced noisy dynamics, commonly known as quantum back-action. Finally, applying feedback which is dependent on the noisy measurement current adds further noise into the dynamics.  Consequently, a variety of feedback control schemes that work well for classical systems may not for the analogous quantum counterparts~\cite{Jacobs2006, Zhang2017,wiseman_milburn_book}.

In recent years, machine learning (ML) has rapidly gained interest, leading to numerous technological advancements in machine vision, voice recognition, natural language processing, automatic handwriting recognition, gaming, and engineering and robotics, to name a few.~\cite{Goodfellow2016} Various ML models, broadly fall into three categories: supervised learning, unsupervised learning and reinforcement learning (RL)~\cite{Goodfellow2016, Mnih2015, Sutton2018}. For supervised/unsupervised methods, the ML model is provided with enough labeled/unlabeled datasets to be trained on, which it uses for discovering the predictive hidden features in the system of interest. On the other hand, RL approaches the problem differently and is not pre-trained with any external data explicitly, but learns in real-time based on rewards. Indeed, RL is regarded as the most effective way to benefit from the creativity of machines, where it collects experiences by performing random experiments on the system (known as the environment in RL literature), learning by trial and error. RL, specially in combination with deep neural networks, abbreviated as DRL, is poised to revolutionize the field of artificial intelligence (AI), particularly with the emergence of autonomous systems which process, in real time, stimuli from real world environments~\cite{Li2019}. 

There have already been several important applications of ML in different areas of physics, such as in statistical mechanics, many-body systems, fluid dynamics, and quantum mechanics~\cite{Cirac2019, Brunton2020, Carrasquilla2017,  Dunjko2018, Mehta2019}. Most of these applications are supervised in nature, e.g., in the quantum domain, these have been applied to solving the many-body systems~\cite{Carleo2017}, in the determination of high-fidelity gates and the optimization of quantum memories by dynamic decoupling~\cite{August2017}, quantum error corrections~\cite{Baireuther2018, Torlai2017, Krastanov2017}, quantum state tomography~\cite{Torlai2018, Neugebauer2020, Lohani2020, Ahmed2020Aug, palmieri_experimental_2020}, classification and reconstruction of optical quantum states~\cite{Ahmed2020}. Recently, a few applications of DRL in quantum mechanics have also appeared that includes applications in quantum control~\cite{Ueda2020, Niu2019Apr, Zhang2019Oct, Xu2021Apr}, quantum state preparation and engineering~\cite{Wang2019, Wrachtrup2020, Haug2020, Guo2021, Bilkis2020Aug}, state transfer~\cite{Prati2019, Ding2020, Paparelle2020}, and quantum error correction~\cite{Marquardt2018, Nautrup2019Dec}. While the number of works utilising DRL is increasing, a very few consider using continuous measurement outcomes explicitly towards training the DRL agent~\cite{Marquardt2018, Ueda2020, Wrachtrup2020}. 
As experiments often employ such continuous quantum measurement techniques for feedback control, we will consider this type of measurement as a key ingredient of our analysis below.

{While traditionally known optimal control techniques  work very well for linear, unitary and deterministic quantum systems, there is no known generalised method for non-linear and stochastic systems. RL, on the contrary, is {\em agnostic} to the underlying physical model, but attempts to control the dynamics of the system by finding patterns from the data produced by it.}
In this Letter, we  model the quantum evolution of a particle in a double well (DW) subject to continuous measurement at a rate $\Gamma$ of the operator ${x}^2$, whose even parity avoids measurement localisation of the particle's wavefunction to either well~\cite{Jacobs2009}. The DRL agent controls the quantum dynamics via a modulation of the Hamiltonian $H^\prime(t) =  {H} + {F}(t)$, with ${F}(t)=\mathcal A(t)( x  p +  p  x)$, where $ x$ and $ p$ are (dimensionless) canonical operators --  a squeezing operator, whose strength $\mathcal A(t)$, is modulated by the DRL agent. The DRL agent is trained via the continuous measurement current,  while, in real time acts back on the system via ${F}(t)$. We show that the DRL agent can be trained to cool the particle close to the ground state. Interestingly, the cooling efficiency depends on the choice of $\Gamma$, for which there is an optimal value of $\Gamma$ to achieve the best cooling which we identify numerically. 

RL translates a problem at hand into a game-like situation in which an artificial agent (also called the controller), finds a solution to the problem based on a trial-and-error approach~\cite{Sutton2018, Mnih2015},  with no hints or suggestions on how to solve the problem itself. For this purpose, the agent is given a policy (in the case of DRL, it is the neural network itself), which is optimized based on some scalar values (rewards) it receives from the environment (that includes the physics of the problem  and the reward estimation function based on the observables) for each decision (action) made by it.  By harnessing the power of search, coupled with many trials, the RL will gain experience from thousands of instances executed sequentially or in parallel in a sufficiently powerful computing infrastructure. After sufficient training,  the agent can become skilled enough to have sophisticated tactics and superhuman abilities as was phenomenally demonstrated by Google's AlphaGO \cite{silver_mastering_2016, silver_mastering_2017}.  {To give a perspective on the applicability of RL in physics and the kinds of tasks it can solve, we provide a short demonstration to a problem in elementary mechanics which we include as media file in the supporting information.}

\begin{figure}[t]
    \centering
    \includegraphics[width=1.0\columnwidth]{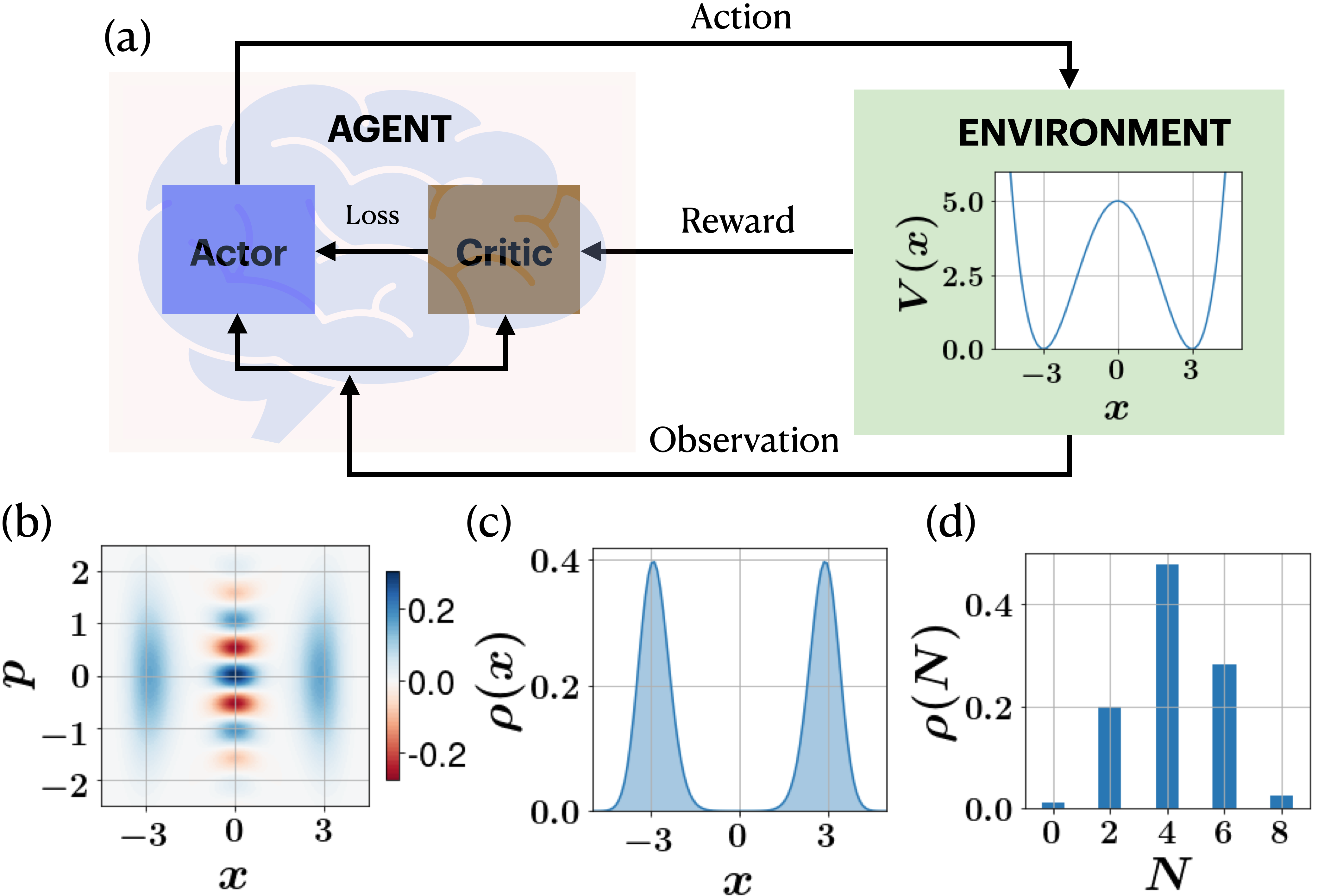}
    \caption{(a) The working of a DRL actor-critic model. The agent consists of two networks, actor and critic, where the actor decides the action to be applied on the environment based on the suggestion made by the critic network that computes the value function based on the reward and state obtained from the DRL environment, that includes the physics of the problem -- the quantum dynamics/state of particle moving in a DW and modulation function to alter quantum dynamics and the reward estimation function based on the observables (measurement results). (b) the Wigner function distribution for the ground state of the DW, and the corresponding probability distribution on position (c) and Fock-number basis (d). The ground state of the DW is an even parity state.}
    \label{fig:model}
\end{figure}

It is possible to implement the DRL agent according to two distinct frameworks, a policy-based  or  value-based framework~\cite{Sutton2018}. In policy-based frameworks, the policy parameters---the weights and biases of the neural network---are optimized directly based on the rewards it receives and informs its future actions on the environment (see Fig.~\ref{fig:model}(a)). Value-based methods on the other hand, optimize the expected future return of a given value function, and deduce the policy from it~\cite{Mnih2015}. It is possible to achieve the best of both these worlds by combining these two approaches in a meaningful way, known as Actor-Critic (AC) methods~\cite{Sutton2018}. Here the actor is the policy which is being optimised and the critic is the value function which is being learned. The actor network can be modeled using various policy-based approaches such as Vanilla Policy Gradient~\cite{Sutton2018}, Trust Region Policy Optimization (TRPO)~\cite{trpo_paper}, or the more recent Proximal Policy Optimization (PPO)~\cite{ppo_paper}. In our work, we used PPO in combination with Advantage Actor-Critic (A2C)~\cite{a2c_paper} as a DRL agent. In the PPO scheme, it optimizes a clipped surrogate objective/loss function given by,
\begin{equation}
\mathcal{L}(\theta) =   \hat {\mathbb{E}}_t \left[ \min \left( r_t(\theta) \hat A_t, \text{clip}\left( r_t(\theta), 1- \epsilon, 1 + \epsilon \right)  \right)\right],
\end{equation}
where, $r_t(\theta) = \frac{\pi_\theta (a_t|s_t)}{\pi_{\theta_{old} (a_t|s_t)}}$ is the probability ratio between current and old stochastic policies, so $r(\theta_{old})=1$. Furthermore, $\hat{\mathbb E}_t$ is the empirical average over a finite batch of samples and $\hat{A}_t$ is an estimator of the advantage function at timestep $t$, obtained from the critic, and calculated as the difference between the Q-value for action $a_t$ at the state $s_t$ and its average value, $V$: $\hat{A_t}(s_t|a_t) =Q(s_t|a_t) - V(s_t)$.  Clipping the ratio within a bound specified by $\epsilon = 0.2$ ensures that the policy is not updated too much. The A2C framework allows synchronous training of multiple parallel worker environments simultaneously, which enables faster training. A more detailed theory can be found in the Supplementary Materials. A depiction of the DRL model employed in this study is shown in Fig.~\ref{fig:model}(a). 

In this article, we will work with dimensionless position and momentum denoted by $(x,p)$. The relationship to the physical position and momentum variables is $x=Q/Q_0, p=P/P_0$ where $Q_0,P_0$ are suitable scales for position and momentum. As the canonical commutation relations are $[\hat{Q},\hat{P}]= i\hbar$ ,thus $[\hat{x}, \hat{p}]=i\bar{k}$,
where the dimensionless Planck's constant is defined by $\bar{k} =\hbar/(Q_0P_0)$. The DW potential we consider is formed along the $x$ axis by the Hamiltonian of a particle
\begin{align}
    H = \frac{p^{2}}{2}+ \frac{h}{b^4}\left( \left(x-a\right)^2 - b^2 \right)^2,
    \label{eq:dw}
\end{align}
where $b$ gives the location of the well's minima, $h$ is the height of the barrier between the wells, and $a$ is the offset along $x$. The ground state of this potential is a type of Schr\"odinger cat state due to the even parity symmetry of $H$ in both $x$ and $p$.
The ground state can be depicted by the Wigner function $\mathcal{W}(x, p)$, which is shown in Fig.~\ref{fig:model}(b).
The probability distribution along the $x$-axis i.e.~$\rho(x) = \int \mathcal{W}(x,p) dp $ is shown in Fig.~\ref{fig:model}(c).  Furthermore, the ground state has even parity symmetry while the first excited state has odd parity ~\cite{Jelic2012}. Hereafter, we will set the parameters $a=0$, $b=3.0$ and $h=5$, which sets the potential to be symmetric around the origin at $x=0$.  It is worthy to note that the DW potential can now be engineered in laboratories such as in superconducting circuits, Bose-Einstein condensates and magneto-optical setups~\cite{Abdi2016}.

To provide data to the agent we consider that the quantum system is subject to a continuous measurement process and these measurement results are provided to the agent in real time. This continuous measurement also induces back-action on the quantum system and noise on both the conditioned quantum dynamics and also on the observed measurement data. We can describe the dynamical evolution of the conditioned density operator $\rho_c(t)$, conditioned on a stochastic measurement record to time $t$ via a quantum stochastic master equation~\cite{wiseman_milburn_book} given by,
\begin{align}
d\rho_c(t) =  - i[H,\rho_c]dt 
+\mathcal D[A]\rho_c\,dt  + \mathcal{H}[A]\rho_c  \, dW(t),
\label{eq:qsde}
\end{align}
where $A$ is a hermitian observable operator under continuous measurement (known as the measurement operator), and $\mathcal{H}[A]$ and $\mathcal D[A]$ are super-operators given by,
\begin{align}
\mathcal{H}[A]\rho_c(t) = \left[\{  A, \rho_c(t)\} - \mathrm{tr}\left(\{A,\rho_c\} \right) \right]\rho(t) \\
\mathcal D[A] \rho_c(t) = \frac{1}{2} \left[2  A \rho_c(t) A^{+} - \{\rho_c(t), A^{+} A\} \right],
\end{align}
where $\{\cdot, \cdot\}$, denotes the anti-commutator. Furthermore, $dW(t)$ in Eq.~\ref{eq:qsde} represents a Wiener increment. It has mean zero and variance equal to $dt$. The measurement result current, $I(t)$, are described by a classical stochastic process that satisfies an Ito stochastic differential equation (see supplementary information)
\begin{align}
I(t)dt = 
\gamma g\left (\langle {A}(t)\rangle_c dt + \frac{1}{\sqrt{4\Gamma}} dW(t)\right ),
\label{eq:homodyne_current}
\end{align}
where $g$ is a gain with inverse units to that of the measurement operator $A$, meaning $I(t)$ has the units of frequency. For the context of the present work, we have $A=\sqrt\Gamma x^2$ and where $\Gamma$  is the measurement rate and quantifies the quality of the measurement (see supplementary information).  As we have fixed units so that $x,p$ are dimensionless we can set $\gamma g=1$. 

\begin{figure}[!hbt]
    \centering
    \includegraphics[width=1.0\linewidth]{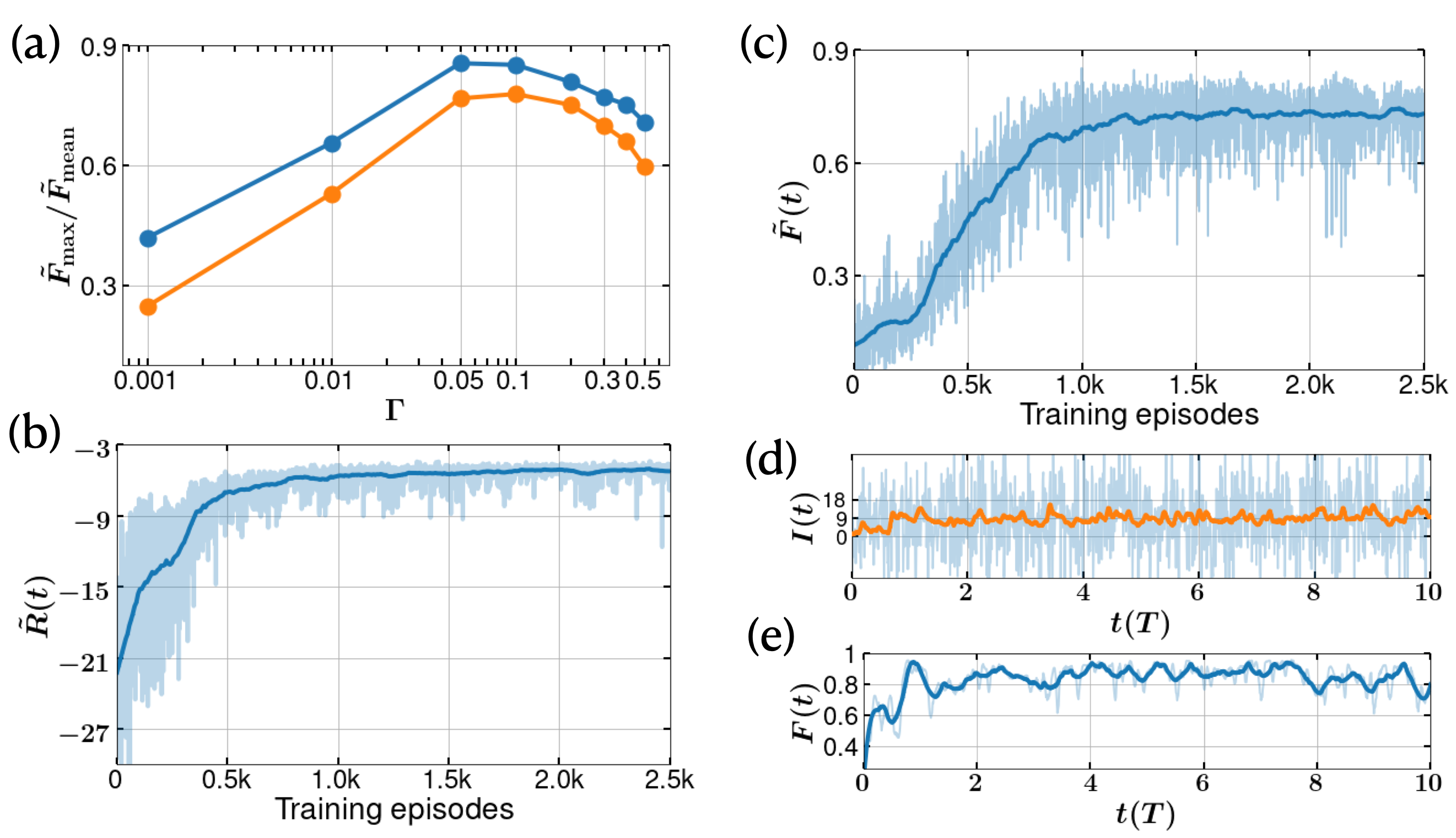}
    \caption{(a) Effectiveness of the agent's learning process as a function of measurement strength, $\Gamma$, indicating the existence of a ``sweet-spot'' around $\Gamma\sim 0.1$. We plot  the maximum ({shown in blue lines}) and mean ({shown in brown lines}) episodic $\tilde{F}$, of 10 successive deterministic episodes where $\tilde{F}$ represents the mean fidelity over an episode using trained agents, (b) The episodic time evolution of the mean reward, $\tilde R(t)$  during the training of the agent when $\Gamma=0.1$ [light(dark) blue includes (average) noise], (c) episodic mean fidelity $\tilde{F}$, of the instantaneous $\rho(t)$  with the ground state of the DW, (d) measurement current, $I(t)$, illustrating that a trained agent is able to keep the wavefunction near the well minima (the conditional average, $\langle x_c^2\rangle$  is shown in brown) and (e) the corresponding variation of fidelity for a trained episode, {for a deterministic episode of the trained agent. Similar performance can be obtained using fidelity as the reward function (see supporting information)}.}
    \label{fig:learning_process}
\end{figure}
To cool the system to the ground state (which is a `cat' state) via continuous measurement, it is important to choose the stochastic operator, $A$ in Eq.~\ref{eq:qsde} as $\sqrt{\Gamma} x^2$ instead of $\sqrt{\Gamma} x$, as the latter would collapse the state to either of the two minima of the DW~\cite{Jacobs2009}.  At each interaction, the agent adds a squeezing term $F(t)\equiv \mathcal A(t) (xp+px)$ to the Hamiltonian (Eq.~\ref{eq:dw}), attempting to adjust the values of $A(t)$ in the continuous range  $\mathcal A(t)\in [-5, 5]$, to maximize the reward. {The choice of such a feedback is motivated by the physics of the problem, which we explain in detail in the supporting information, backed by an analysis using Bayesian control driven by the conditional mean of the measurement record following the method of Stockton et al.,~\cite{Stockton2004Aug}. }It is possible to implement $xp-$type Hamiltonian terms via motion in a magnetic field ~\cite{Twamley2018}. In each episode of the training process the DRL interacts with the environment $1000$ times,  in intervals of $\delta t = 0.01$, and each time applies an action to the environment. {Further detail of the implementation and other technicalities of the DRL can be found in the supporting information.}

The amplitude of the measurement noise depends on two parameters - (a) the measurement strength, $\Gamma$ and (b) the measurement time, $\delta t$. Since the Wiener noise in Eq.~\ref{eq:homodyne_current} is a Gaussian with variance {$\delta t$},  the noise term in the measurement current ${\cal I}(t)$, scales at least as $1/\sqrt{4\Gamma\delta t}$. Because of this, one might expect the DRL to learn more efficiently for larger values of $\Gamma$, however, this is not the case. This makes it critical to choose an optimal value for $\Gamma$ along with the measurement time $\delta t$. For a choice of $\delta t=0.01$, we observe that optimal learning occurs near $\Gamma\sim 0.1$, and worsens for other values of $\Gamma$. Similar effects can be observed in Markovian measurement-based direct feedback which we discuss in the supplementary information. Larger $\Gamma$ values result in an increase in noise, while smaller $\Gamma$ values return a very low signal to noise ratio in the continuous measurement process. The agent tends to learn most efficiently when the dynamics fluctuate in a limited domain around the DW minima.  The effectiveness of the agent learning is shown in Fig.~\ref{fig:learning_process}(a), in terms  mean and maximum fidelities from 10 successive deterministic episodes  of the agents trained on the measurement current, $I(t)$.

An important ingredient in the DRL  is a suitable choice for the reward function.  Many research studies have previously used fidelity or energy as the reward function. However, such a function is not practically available in experiments. Instead we propose a measurement based reward function $R(t) = - \left|{I}(t)/(\gamma g)-3^2\right|$, where ${I}(t)$ is the  measurement current (Eq.~\ref{eq:homodyne_current}). This function obtains its maximum reward of $0$, when $\langle x_c^2\rangle =3^2$, at the well minima positions. The learning process of the agent is shown in Fig.~\ref{fig:learning_process}(b), along with the   fidelity of the instantaneous state of the system  with respect to the DW ground state in  Fig.~\ref{fig:learning_process}(c). Although such a fidelity is not possible to evaluate experimentally in real-time, we present this as a check of the learning process. In a given episode the trained agent is able to adapt the feedback in such a way that the particle oscillates near the minima of the DW, as shown in Fig.~\ref{fig:learning_process}(d) (for $\gamma g=1$). The corresponding variation of  fidelity for the episode is shown in Fig.~\ref{fig:learning_process}(e). It is worthy to note that with a different choice of the reward function it might be possible to obtain better and more stable learning, see supporting information for details.

At the beginning of each episode, the environment must be reset to an initial state (initial density matrix, $\rho(0)$), which is needed to start the stochastic master equation solver. We have found that the choice of $\rho(0)$ plays a crucial role in determining the total reward that can typically be achieved by the agent. When $\rho(0)$ is a thermal or coherent state, the DRL converges to an average fidelity of about 60\%. However, if we use a small cat state or the ground state of the DW itself, the agent is able to achieve a mean trained fidelity of over 80\% with noisy measurement data.  The parity of the initial state is crucial as the stochastic process of continuous measurement and feedback we have chosen is parity conserving. The target ground state of the DW however has even parity and so choosing an initial state with a component of odd-parity will lower the ultimate fidelity achievable. We achieve similar high performance if we start with a thermal state projected on to even parity, as done for Fig.~\ref{fig:learning_process}. {The explicit comparison of the performance of the trained agent with an untrained one is demonstrated in the supporting media.}

{We benchmark our results against the state-based Bayesian feedback protocol (where the feedback is based on an estimate of the state) as proposed by Doherty et al.~\cite{Doherty1999Oct, Stockton2004Aug}. In the context of the present work, the protocol can be simplified to provide feedback of the form $\mathcal{F}(t)= -(\langle {x}_c^2\rangle(t) - 3^2) \times (xp + px)$ where $\langle x_c^2 \rangle (t)$ denotes the conditional mean of the observable $x^2$. We find that this Bayesian control achieves a mean fidelity of $\sim 85\%$. However, $\langle x_c^2 \rangle (t)$ is not a quantity directly accessible in real experiments. When the Bayesian feedback is instead driven by the noisy measurement current $I(t)$ (which is available in experiments), we find that Bayesian feedback demonstrates almost no control over the dynamics.  Numerical simulations with 1000 copies of the system (ensemble), evolving under a given feedback based on the mean of the measurement currents during each time step, yields an average fidelity of $\sim 42\%$. This is considerably worse than the performance of the DRL agent. A more detailed discussion can be found in the supporting information. }

{We have found that the DRL shows a robust control when the measurement efficiency, $\eta > 50\%$, as shown in Fig.~S5(a) of the supporting information implying that the the DRL-agent is able to find patterns in the underlying dynamics even when the stochasticity is significantly increased. Similarly, the DRL shows no significant drop of fidelity under additional dephasing of the form $\sim\sqrt \gamma a^\dagger a$, but this fidelity does drop for damping  $\sim\sqrt \gamma a$, where $\gamma$ is the decoherence rate (see supporting information). 

On the computational side, the challenge for DRL control is the significant computational expense, e.g.  3-4 days of simulation time, even with fast computational CPUs, the bottleneck being the slow stochastic solver routines.  For the RL side, we find that RL algorithms does not scale up linearly with the number of parallel processes, and thus improved parallelization in such RL computations could be worthwhile. We expect improved performance if the agent is made to learn in a dynamic combination of supervised (under a supervised setting a ML agent can learn more effectively from less data points, but is not reward based), and reinforcement learning (which is reward based and hence useful for feedback control), as done for image recognition~\cite{Kangin2018Jul,Senft2017Nov, Wang2018Jul}. It is possible to reshape the data to images of actions and measurement records which would enable the usage of convolution neural networks (CNN) in GPUs/TPUs with multi-core support and  utilise different image compression techniques, e.g.,  deep compressed sensing technique, proposed recently by researchers from DeepMind~\cite{wu2019deep}. From the physics side, use of proper filters (as normally done in experiments) to filter the noisy signals prior to inputting into the DRL is expected to be crucial.  In addition, the use of better reward estimation, such as combining constraints on current, fidelity (using tomography), and energy, is expected to be useful for further improvement.  An even further innovation would be to use RL in combination with various optimal and Bayesian control protocols, as recently explored in applications outside of quantum mechanics~\cite{Shashua2020Feb, Carron2016Dec, Krastanov2017}}

In conclusion, we have demonstrated the usefulness of  deep reinforcement learning to tailor the non-trivial feedback parameters in  a non-linear system to engineer evolution towards the ground state. We found that the artificial agent can discover novel strategies solely based on measurement records to engineer high-fidelity `cat states' for the quantum double well. 
\section{acknowledgments}
\begin{acknowledgments}
	The authors thank the super-computing facilities provided by the Okinawa Institute of Science and Technology (OIST) Graduate University and financial support. GJM and MK acknowledge the support of the Australian Research Council Centre of Excellence for Engineered Quantum Systems CE170100009. 
\end{acknowledgments}

\newpage
\widetext

\author{Sangkha Borah}
\email{sangkha.borah@oist.jp}
\affiliation{Okinawa Institute of Science and Technology Graduate University, Onna-son, Okinawa 904-0495, Japan}
\author{Bijita Sarma}
\affiliation{Okinawa Institute of Science and Technology Graduate University, Onna-son, Okinawa 904-0495, Japan}
\author{Michael Kewming}
\affiliation{Centre for Engineered Quantum Systems, School of Mathematics and Physics, University of Queensland, QLD 4072 Australia}
\author{Gerard J. Milburn}
\affiliation{Centre for Engineered Quantum Systems, School of Mathematics and Physics, University of Queensland, QLD 4072 Australia}
\author{Jason Twamley}
\affiliation{Okinawa Institute of Science and Technology Graduate University, Onna-son, Okinawa 904-0495, Japan}

\def\theequation{S\arabic{equation}}
\renewcommand{\thepage}{S\arabic{page}} 
\renewcommand{\thesection}{S\arabic{section}}  
\renewcommand{\thetable}{S\arabic{table}}  
\renewcommand{\thefigure}{S\arabic{figure}}
\setcounter{figure}{0}
\setcounter{table}{0}
\setcounter{section}{0}
\setcounter{subsection}{0}
\setcounter{page}{1}
\begin{center}
	\textbf{\large Measurement Based Feedback Quantum Control With Deep Reinforcement Learning for Double-well Non-linear Potential: Supporting Information}
\end{center}
\maketitle

\section{A quick introduction to reinforcement learning}
Essentially there are three kinds of machine learning, namely supervised learning, unsupervised learning and reinforcement learning~\cite{Goodfellow2016}. In supervised learning, the ML model is provided with sufficient input and output data (labeled data), which are used to train the model to discover the hidden pattern/features in the datasets. For example, the dataset may consist of RGB images of different fruits, e.g., of apples, oranges, and bananas, with proper labels assigned by their names. The ML model can read the pixels of those images and correlate to the labels in the datasets by training on them. After training on enough datasets, the ML model is expected to acquire the necessary wisdom to predict from a new image (on which no training was done) whether it is an apple or not.  On the other hand, in unsupervised learning, we do not provide labels to the datasets. After training, the model is able to make a classification of those fruits based on the hidden features in the pixel data of the images, and eventually acquire the expertise to classify new test images of fruits again, whether it is an apple, orange or banana or not. 
\begin{figure*}[h]
    \centering
    \includegraphics[width=0.9\textwidth]{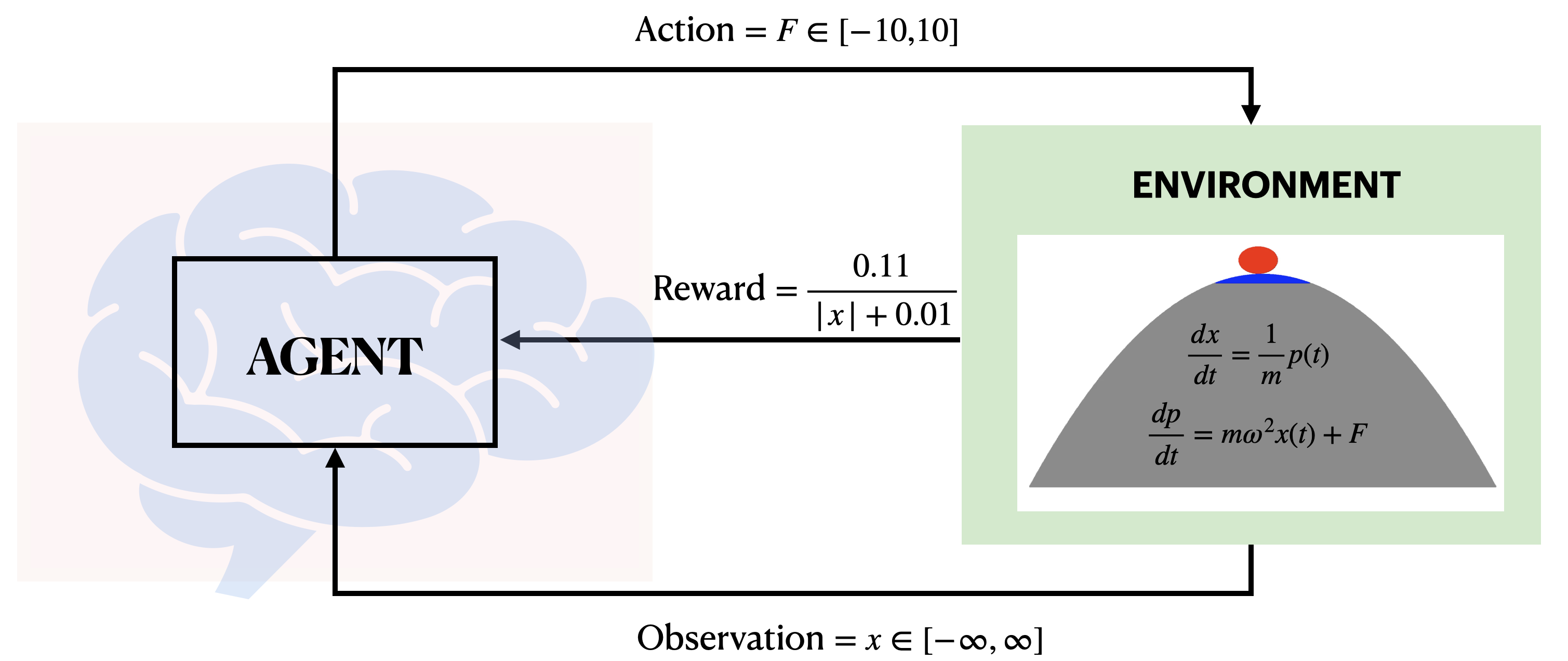}
    \caption{The workflow of reinforcement learning, with the environment as the classical upside-down harmonic oscillator. At a given time the agent exerts an action (a discrete force), that modifies the dynamics of the environment (leads to a direct change in the equations of motion). The new position $x$ of the oscillator's position at a short time later is the observation (stimuli), received by the agent, along with a reward (chosen as $r(t)=0.11/(|x(t)|+0.01)$. This reward is maximised at $x=0$ and gradually decreases away from the origin). This reward value and associated observation stimuli are the information upon which the agent modifies its parameters and decides a new action. The loop is repeated for enough episodes until the agent receives a saturated net maximum reward for its actions taken per episode.}
    \label{fig:sup_rl}
\end{figure*}

Reinforcement learning (RL)~\cite{Sutton2018} is the third kind of ML, the basic workflow of which is shown in Fig.~\ref{fig:sup_rl}. It comprises of two main characters - the agent and the environment. The environment is the agent's world in which it lives and interacts with. The agent is the brain of the RL and comprises of routines to function approximators, such as the artificial neural networks. In the case when the agent uses deep neural networks as function approximators (neural networks are non-linear function approximators), the RL is known as deep reinforcement learning (DRL). The RL-agent periodically interacts with the RL-environment in real time, in which it applies some changes to the dynamics of the system by changing some dependent variables. The changes it applies to the environment are called the actions. In return, the agent gets back full/partial information of the changes in the dynamics of the environment (known as the observation or the state of the system) along with a scalar value known as the reward signal. The reward function returns a scalar value that gives an estimation of the likelihood of the applied action bringing to desirable changes in the dynamics or not. The idea behind choosing a  reward function is that rewarding or punishing an agent for its behaviour (actions applied) makes it more  likely to repeat/forgo that behaviour in future. For that, the agent is set the goal to maximize the cumulative rewards (known as return) by changing its actions within a certain number of interactions in real time, known as an episode (a play in games). At the beginning of such a RL experiment, the agent applies the actions randomly, and later updates the parameters of the agents function approximators (the weights and biases of the neural networks in the case of DRL) based on which the next actions are chosen. In reality, the agent-environment loop has to be computed for several hundreds/thousands of episodes depending on the complexity of the problem of interest. For instance, for stochastic environments with noisy data, the learning process of the RL-agent is slower than compared to its deterministic counterpart. 

To understand it better, we here try to demonstrate the working of DRL with a fundamental classical mechanics problem, shown in Fig.~\ref{fig:sup_rl}. The environment is an upside-down harmonic oscillator with a completely frictionless surface, such that a ball can not be held fixed on top of it without any external forces. The task of the agent would be to apply engineered amount of forces back and forth to keep it on top, namely in the blue region shown in the figure. The reward function is chosen such that it gets the maximum reward at $x=0$ and gradually smaller ones as we go away from the origin. This must be programmed together with the solver of the equations of motion in the RL environment. The agent, composed of a few layers of neural networks (hence known as deep reinforcement learning, DRL) can be  trained for few thousand steps before it acquires superhuman or human-like skills to adapt the external forces to keep it stable within the blue region. A video demonstration of the same can be obtained in the supporting media or the GitHub link ~\cite{ml-iho-demo}.

The rule based on which the agent decides the action to be applied on the environment is known as its policy. In the case of DRL the rule roughly represents the neural network function approximators, as the rule is implicitly determined by some non-intuitive choices of the parameters (weights and biases) of the neural networks.  A policy can be deterministic, in which case it is usually denoted by $a_t = \mu_\theta(s_t)$ or stochastic, which is denoted by $a_t \sim \pi_\theta(\cdot |s_t)$, where $a_t$ is the action chosen by the agent  as a function of the state $s_t$ at time $t$. The dot ($\cdot$) in the case of stochastic policy ($a_t \sim \pi_\theta(\cdot |s_t)$) represents the conditioning on $s_t$ (probabilistic), unlike the case of deterministic policy $a_t = \mu_\theta(s_t)$, for which the action is a deterministic function that maps the state $s_t$ to the the policy $\mu$ at time $t$. A deterministic policy is usually chosen only for evaluating the performance of an agent after being trained. 

The reward function $R$ is one of the most critical parts of RL, which is crucial for the agent's learning. In its most general form, it is a function of the current state, the action applied and the next state of the environment, $r_t = R(s_t, a_t, s_{t+1}$, where $r_t$ is the scalar reward signal at time $t$. The cumulative reward is defined as, \begin{equation}
    R(\tau) = \sum_{t=0}^{T} r_t,
    \label{eq:s_return}
\end{equation}
where $\tau = (s_0, a_0, s_1, s_2 ...s_t, a_t .. )$ represents the  sequences of state and action over an episode known as a trajectory, and $T$ represents the horizon, giving the extent to which the summation is done. In this case, the return is known as finite-horizon un-discounted return. Often the choice is to use infinite horizon limit with a discount factor $\gamma \in (0, 1)$, known as infinite-horizon discounted return: 
\begin{equation}
    R(\tau) = \sum_{t=0}^{\infty} \gamma^t r_t,
    \label{eq:s_return_discounted}
\end{equation}
which essentially signifies the fact that rewards now is better than rewards later. The choice of the discounted rewards also makes it mathematically convenient. 

The task of the agent is to maximize the expected return  $J(\pi) = {\hat {\mathbb{E}}_t} \left[R(\tau) \right]$, where ${\hat {\mathbb{E}}_t}$ is the empirical average over a batch of samples, for which the policy is optimized for the optimal policy $\pi^*$: $\pi^* = \text{arg} \max_{\pi} J(\pi)$. The expected return for choosing a given state or state-action pair is known as the value function. The expected return for the starting in a state $s_0 = s$ and following the policy afterwards gives the on-policy value function, given by,
\begin{eqnarray}
V^\pi (s) = \underset{\tau \sim \pi}{{\hat {\mathbb{E}}_t}} \left[R(\tau) |s_0 = s \right].
\end{eqnarray}
On the other hand, the expected return for the starting in a state $s_0 = s$ and taking an action $a_0 = a$ and following the policy afterwards gives the on-policy action-value function, given by,
\begin{eqnarray}
Q^\pi (s, a) = \underset{\tau \sim \pi} {\hat {\mathbb{E}}_t} \left[ R(\tau) |s_0 = s, a_0 = a \right].
\end{eqnarray}
The value functions obey the Bellman equations, which can be solved self-consistently, for example, for action-value function, it is given by,
\begin{eqnarray}
Q^\pi (s, a) =   {{\hat {\mathbb{E}}_t}} \left[ r(s, a)  + \gamma . \max\limits_{a^\prime} Q_{\theta} (s^\prime, a^\prime) \right].
\label{eq:sl_q_leaning}
\end{eqnarray}
Sometimes, yet another function that combines both the types of values function is defined, given by, 
\begin{eqnarray}
\hat A^{\pi}(s, a) = Q^\pi (s, a) - V^\pi (s).
\end{eqnarray}
This is called the advantage function and gives a measure of advantage of taking an specific action $a$ in state $s$ over the random selection of it following $\pi(\cdot|s)$. 

In recent years, several different RL algorithms have been proposed, among which the leading contenders are  the deep Q-learning~\cite{Mnih2015}, vanilla policy gradient methods~\cite{Sutton2018}, and trust region natural policy gradient methods~\cite{trpo_paper, ppo_paper}. In the modern scenario, a successful RL method is expected to be scalable to large models, parallel implementations, data efficiency, and robustness, and should be successful on a variety of problems without much of hyper-parameter tuning. While Q-learning methods are known for their efficiency in numerous discrete action-based environments, such as various Atari games~\cite{silver_mastering_2016, silver_mastering_2017}, they fail on many simple problems, and  may suffer from the lack of guarantees for an accurate value function, the intractability problem resulting from uncertain (stochastic) state information as well as the complexity arising from continuous states and actions~\cite{ppo_paper}. Vanilla policy gradient methods~\cite{Sutton2018} are known for better convergence properties, are effective in high-dimensional or continuous action spaces, and can learn stochastic policies. However, they often converge to a local minimum rather than a global one, and have poor data efficiency and robustness, often leading to high variance in policy updates. The trust region policy optimization (TRPO)~\cite{trpo_paper} is relatively complicated, and is not compatible with architectures that include noise such as dropout or parameter sharing between the policy and value function, or with auxiliary tasks. Proximal policy optimization (PPO)~\cite{ppo_paper} is the latest inclusion in the list, which is shown to yield similar or better performance and robustness as TRPO, with much easier implementation and parallelization, and faster optimization of policy parameters with stochastic gradient methods, and is naturally  compatible with noisy environments.    
\begin{figure*}[!]
    \centering
    \includegraphics[width=0.9\textwidth]{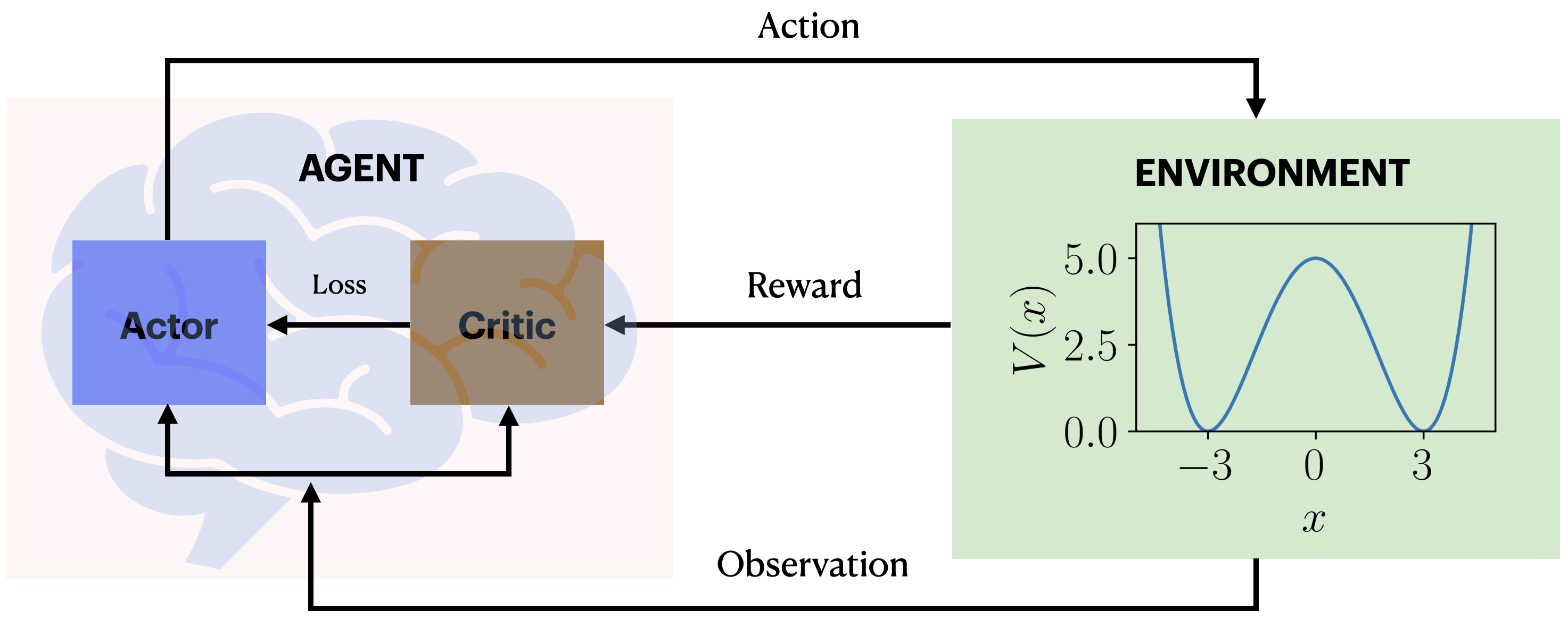}
    \caption{The workflow of an actor-critic reinforcement learning agent. The actor is based on a policy network that is responsible for any action taken on the environment, while the critic network receives the reward from the environment and makes useful suggestion to the actor by updating the parameters of the action value function.}
    \label{fig:s_a2c}
\end{figure*}

Policy gradient methods work by optimizing the following objective (loss) function:
\begin{equation}
\mathrm {L}^{\rm PG}(\theta) = \hat {\mathbb{E}}_t \left[ \log \pi_\theta(a_\theta|s_t)\hat A_t\right],
\end{equation}
where, $\pi_\theta$ is a stochastic policy, and $\hat{A}_t$ is an estimator of the advantage function at timestep t. $\hat{\mathbb E}_t$ is the empirical average over a finite batch of samples. In TRPO methods, the objective function (the “surrogate” objective) contains the probability ratio $r(t)$ of the new ($\pi_\theta$) and old policies ($\pi_{\theta_{old}}$) instead of $\log \pi_\theta$:
\begin{equation}
L^{\rm TRPO}(\theta) = \hat {\mathbb{E}}_t \left[  \frac{\pi_\theta(a_t|s_t)}{\pi_{\theta_{old}}(a_t|s_t)} \hat A_t\right] =  \hat {\mathbb{E}}_t \left[  r_t(\theta) \hat A_t\right],
\end{equation}
under the constraints that the average KL-divergence between new and old policies are less than a small cutoff, $\delta$:
\begin{equation}
    \hat {\mathbb E}_t \left[\mathrm{KL}\left[ \pi_{\theta_{old}}(\cdot|s_t),\,\, \pi_{\theta}(\cdot|s_t)\right] \right] \ll \delta.
\end{equation}
Without the constraint, maximizing the objective would lead to an excessively large policy update~\cite{trpo_paper}.  PPO modifies the above constraint by clipping the policy update, with a hyper-parameter $\epsilon \sim 0.2$, and optimizing the objective function with stochastic gradient method~\cite{ppo_paper}:
\begin{equation}
L^{\rm PPO}(\theta) =   \hat {\mathbb{E}}_t \left[ \min \left( r_t(\theta) \hat A_t, \text{clip}\left( r_t(\theta), 1- \epsilon, 1 + \epsilon \right)  \right)\right].
\end{equation}

Q-learning, on the contrary, optimizes value function $Q_\theta(s, a)$ using the Bellman equation, as given in Eq.~\ref{eq:sl_q_leaning}. Lately, different algorithms that combines policy gradient methods with Q-learning methods have been proposed and have been found to sidestep many of the drawbacks of the traditionally used approaches. These are known as the actor-critic methods, in which there are two policy networks, actor and critic. While the actor, based on a policy network, takes an action on the environment, the critic network receives the reward from the environment and makes useful suggestion to the actor by updating the parameters of the action value function.  In this paper, we have used a DRL agent that combines PPO with advantage actor-critic (A2C) approach~\cite{a2c_paper} , where the advantage function is  is calculated as the difference between the Q-value for action $a_t$ at the state $s_t$ and the average value of the state: $\hat{A_t}(s_t|a_t) = Q(s_t|a_t) - V(s_t)$, discussed above.  The A2C framework allows synchronous training of multiple parallel worker environments simultaneously, which enables faster training. The basic workflow of A2C is shown in Fig.~\ref{fig:s_a2c}.

\section{Weak continuous measurement model}
We have used a weak continuous measurement of $\hat{x}^2$ described using a master equation for the system density operator. In this section we derive this equation using a continuous limit for a sequence of weak measurements at random times.

A single  weak measurement of a Hermitian operator $\hat{A}$ is described using an operation on the state of the system that gives the post-measurement state conditioned on a measurement result $z\in {\mathrm  R}$  
\begin{equation}
   \rho_{|z} = [P(z)]^{-1}\hat{\Upsilon}(z)\,\rho\, \hat{\Upsilon}^\dagger(z),
\end{equation}
where $P(z) = {\rm Tr}[\hat{\Upsilon}^\dagger(z)\hat{\Upsilon}(z)\rho] $ is the probability density of the measurement result and  
\begin{equation}
   \hat{\Upsilon}(z)=(2\pi\sigma)^{-1/4}\exp\left[-\frac{(z-g\hat{A})^2}{4\sigma}\right],
\end{equation}
with $[\hat{A},\hat{\Upsilon}(z)]=0$, and $g$ is the gain in inverse units to $\hat{A}$. The measurement result, $z$, is a classical random variable conditioned on the state of the measured system. The mean and variance of $z$ are given by
\begin{eqnarray}
{\cal E}[z] & = & g\langle \hat{A}\rangle  \\
{\cal V}[z]  & = & g^2\langle \Delta \hat{A}^2\rangle +\sigma
\end{eqnarray}
where $\Delta \hat{A}= \hat{A}-\langle \hat{A}\rangle $ with the quantum averages defined by $\langle \hat{A}\rangle ={\rm tr}[\hat{A}\rho]$. It is clear that $\sigma$ can be interpreted as the measurement error over and above any intrinsic quantum uncertainty. The {\em unconditional } state of the system post measurement is then given by 
\begin{equation}
    \rho'= \int dz P(z) \rho_{|z}= \int dz \hat{\Upsilon}(z)\rho \hat{\Upsilon}^\dagger(z).
\end{equation}

We now move to a weak continuous measurement by assuming each of these single weak measurements occur at random, Poisson distributed times. At each time, the device that performs the measurement is discarded after the measurement result is recorded,  and a new measurement device is supplied - ready for the subsequent measurement. This builds in the Markov condition for the irreversible measurement process. The master equation  describing the unconditional dynamics of the entire system is given by \cite{Milburn1987}, 
\begin{equation}
\dot{\rho} = \gamma \int_{-\infty}^\infty  \ dz \hat{\Upsilon}(z)\rho\hat{\Upsilon}^\dagger(z)-\frac{1}{2}( \hat{\Upsilon}^\dagger(z)\hat{\Upsilon}(z)\rho+\rho  \hat{\Upsilon}^\dagger(z)\hat{\Upsilon}(z))\;\;,
\end{equation}
where $\gamma$ is the rate of measurements.

We now take the limit in which the rate of measurements satisfies $\gamma \delta t \gg 1 $ where $\delta t$ is the time scale of the free dynamics of the system and the uncertainty of the measurements $\sigma \gg  g^2$ such that the ratio $\gamma/\sigma$ is constant. This is the continuous weak measurement limit. The unconditional  master equation then becomes
\begin{equation}
\dot{\rho} = -{\Gamma}[\hat{A},[\hat{A},\rho]],
\end{equation}
with $\Gamma =\gamma g^2/\sigma$. The effect of the double commutator term in this master equation is to destroy coherence in the diagonal basis of $\hat{A}$ and adds diffusive noise to incommensurate variables $\hat{B}$  where $[\hat{A},\hat{B}]\neq 0$. A good measurement requires that $\Gamma$ is large. 

We define an observed {\em classical} stochastic process in the continuum limit as follows. 
Denote a complete record of the measurement results, indexed by the random times of the measurement times, as  $z(t )$. Now define the observed process by the stochastic differential equation
\begin{equation}
dy(t)=z(t)dN(t),
\end{equation}
where $dN(t)^2=dN(t)$ and $\overline{dN(t)}=\gamma dt$ while $z(t)$ is a Gaussian distribution. In the continuum limit this point process can be approximated by a diffusion process as 
\begin{equation}
dy(t) =\gamma g\left (\langle \hat{A}(t)\rangle_c dt+(4\Gamma)^{-1/2} dW(t)\right ),
\label{eq:sl_dy}
\end{equation}
where the subscript $c$ indicates that this average must be computed from the  state conditioned on the entire history of $y(t)$ to time $t$ and $dW(t)$ is the Wiener process. Note  that for a good measurement in which $\Gamma$ is large, the signal-to-noise ratio also becomes large. 

In order to evaluate the conditional averages in the observed process we need a continuous limit of the conditional state at each measurement. The result is the stochastic master equation~\cite{wiseman_milburn_book},
\begin{equation}
d\rho_c= -i[H, \rho_c]dt  + \mathcal{D}[\hat A]dt+\sqrt{2\Gamma}{\cal H}[\hat{A}]\rho_c dW(t),
\label{eq:sl_sde}
\end{equation}
where 
\begin{align}
  \mathcal{D}[A] = -{\Gamma}[\hat{A},[\hat{A},\rho_c]],
\end{align}
and
\begin{equation}
{\cal H}[\hat{A}]\rho_c =\hat{A}\rho_c+\rho_c\hat{A}-{\rm tr}(\hat{A}\rho_c+\rho_c\hat{A}).
\end{equation}
This equation is non linear; not surprising as for a conditional process the future must depend on the entire past history of measurement results. 

\section{Markovian Quantum Feedback}
Let us consider the case when we remove the DRL agent altogether and set the feedback Hamiltonian to be proportional to the measurement current, $dy(t)/dt$ (Eq.~\ref{eq:sl_dy}). The effect of feedback on $\rho_c$ can be introduced as,
\begin{equation}
    d\rho_f = [\exp(\mathcal K dy) - 1] \rho_c,
\end{equation}
where $\mathcal K$ is a super-operator such that $\mathcal K \rho_c = -i[F, \rho_c]$ for some Hermitian feedback action operator $F$. With  Eq.~(\ref{eq:sl_sde}), we get the modified closed-loop stochastic master equation under feedback $F$,
\begin{align}
d\rho_f = \Big\{ -i[H, \rho_f] + \Gamma \mathcal{D}[\hat A]\rho_f - i[F, \hat A\rho_f + \rho_f \hat A] + \mathcal{D}[F] \rho_f\Big\} dt + \mathcal{H}\left[\hat A - iF \right]\rho_f dW.
\end{align}
Averaging over the noise we get the the Wiseman-Milburn unconditional master
equation with Markovian feedback as
\begin{equation}\label{Wiseman-Milburn master equation}
\dot{\rho}=-i\left[H,\rho\right]+\Gamma
\mathcal{D}\left[\hat A \right]\rho-i\sqrt{\Gamma}\left[F, \hat A\rho+\rho
\hat A\right]+\mathcal{D}\left[F\right]\rho. 
\end{equation}
where $\rho = E(\rho_f)$, the average over $\rho_f$.  Looking at the unconditional master equation (\ref{Wiseman-Milburn master equation}), one can identify the feedback term, $-i[F, \hat A\rho + \rho \hat A]$, which modifies the base Hamiltonian evolution generated by $H$. The term $\Gamma{\mathcal D}[\hat A]\rho$, is the decoherence introduced by the continuous measurement of $\hat  A$. However we now observe a new source of decoherence ${\mathcal D}[F]\rho$, introduced by the feedback itself~\cite{Zhang2017}. There is often this competition between the two sources of decoherence that decide an optimal value of $\Gamma$. For small values of $\Gamma$, the measurement current $dy/dt$, is very noisy and the feedback decoherence is high, while the continuous measurement decoherence is low. For large values of $\Gamma$, the feedback decoherence is low while the measurement decoherence is high.  This competition carries over to the case studied in the main manuscript where  the feedback action is determined by a DRL agent. We thus observe an optimal measurement rate, $\Gamma$ for the learning of the DRL agent, as shown in Figs.~2(a) and \ref{fig:supp_fidelity_as_control}(a).

\section{The DRL in the study}
Using the model given in Eq. (2) of the paper, we created a DRL environment (shown in Fig.1(a) using the open-source platform OpenAI-Gym~\cite{OpenaiGym}, in which QuTIP's~\cite{QuTiP} stochastic master equation (SME) solver is used to compute the dynamics, see Eq.~(6). The environment primarily includes two routines-- the quantum dynamics/state of particle moving in a DW and modulation function to alter the quantum dynamics and a reward estimation function based on the observables (measurement results) in a step-wise manner for every interaction made by the agent. Our PPO agent~\cite{ppo_paper} is constructed following the implementations of stable-baselines3 ~\cite{stable-baselines3} in the A2C settings~\cite{a2c_paper}, where both the actor and critic are modelled with a set of fully connected feed-forward networks dimensions $512 \times 256 \times 128$, with the first layer as a shared network between the two.  For modeling the neural networks we have used the open-source ML platform PyTorch~\cite{pytorch}. The input of the network is provided as the mean of  the  measurement currents, $I(t)$ from the last four time-steps (which helps learning faster rather than using the instantaneous current), along with the action returned by the network. In each episode of the training process the DRL interacts with the environment $1000$ times,  in intervals of $\delta t = 0.01$, and each time applies an action to the environment. The agent gathers trajectories as far out as the horizon limits (in our case 4000 steps, i.e., data from 4 episode for each environment) for each environment running in parallel (in our case 8 environments in total), then performs a stochastic gradient descent update of mini-batch size (in our case it is 100) on all the gathered trajectories for the specified number of epochs (in our case 10). We have found that a larger network along with a small learning rate ($10 ^{-5}$) helps the agent keep the learning process stable during longer simulations. 

These simulations require a high truncation of the Hilbert-space dimension (we took $N_\mathrm{trunc}=60$), since the process of measurement often drives the state to high Fock numbers (see supporting media for demonstration or the GitHub link~\cite{ml-dw}). This is especially important for the DRL in the initial stages of learning when it is trying to gather information about the system by performing random actions. In our numerical experiments we chose $N_\mathrm{trunc}=60$,  $\Gamma \leq 0.5$, and used 8 parallel vector environments to speed up the processing. After training within a few hundred  such  parallel environments the DRL learns to apply the actions in such a way that it limits the occupation to low Fock states, avoiding and errors with truncation limits. In addition, we observe that the efficiency of learning depends on the type of agent.  Although traditional DQN  methods based on discrete actions works well with simple conditional mean data, we find that they perform very poorly in the presence of noise. Actor-critic methods, especially the newer PPO agent (an evolution of A2C and TRPO), is better suited for dealing with a stochastic environment (stochastic data) than a purely value-based algorithm like DQN. 

\section{Role of feedback Hamiltonian}
\begin{figure}[h]
    \centering
    \includegraphics[width=1\linewidth]{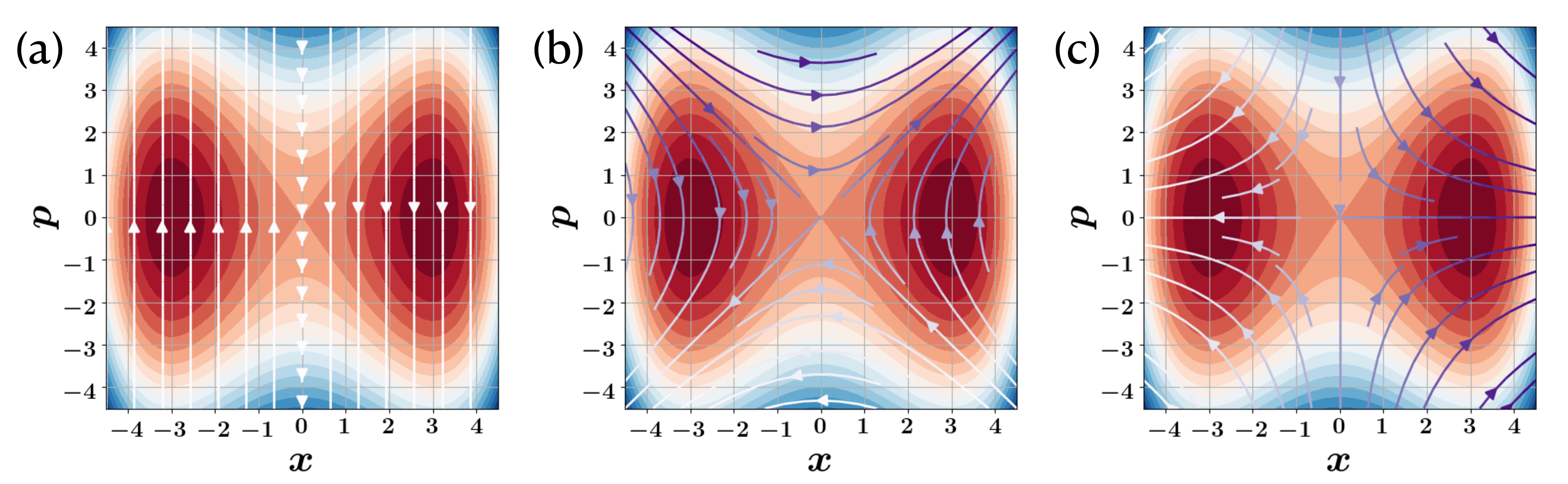}
    \caption{The phase space streamlines of different action feedback Hamiltonians: (a) $x^2$, (b) $p^2 - x^2$ and (c) $xp+px$. The equipotentials of the DW potential are shown as filled contours with minima at $x=\pm 3$.}
    \label{fig:xp_feedback}
\end{figure}
In the main manuscript we have chosen the agent controlled action to be $F(t)\propto (xp+px)$. This action augments the default DW quantum dynamics. It is natural to enquire if there are other, more effective actions that will improve the learning of the DRL process to achieve the goal of cooling the DW to its quantum ground state. Interestingly, we find that this form of action appears to optimally effective and choosing other forms of action results in weaker DRL efficiency. The action operator, $F(t)=A(t)f$, where $f = x  p +  p x$, is the Hermitian squeezing operator. The reason why this action is so effective in learning and controlling the quantum dynamics of the particle in the double-well potential can be understood merely from classical perspectives. 

Consider  the classical streamlines of the feedback action Hamiltonian in the phase space of the double-well potential. The feedback, $f(x, p)$ generates a flow vector  $(V_x,  V_p)$ in classical phase space defined by, $V_x = \{x,f\}$ and $V_p  = \{p,f\}$, 
where $\{,\}$ is the Poisson bracket. This flow is always tangential to the curves of $f=\text{constant}=f(x_k,p_k)$,   where $(x_k, p_k)$ is a phase space point at time $t_k$ just before feedback, and moves the particle along curves of constant $f$. Fig.~\ref{fig:xp_feedback} shows the phase space streamlines for three different possible feedback action Hamiltonians. Feedback in the form of $f=x^2$ is likely to be very ineffective, as indicated by the streamlines in Fig.~\ref{fig:xp_feedback}(a), as the flow is always orthogonal to the $x$-axis. Similarly, the streamlines for $f=p^2 -x^2$ would push the state starting near the $p$ axis further away from the $x$ axis. On the other hand, for the case of $xp+px$, the streamlines push the particle in the direction of the x-axis, which is appropriate for approaching the double-well minima.

Numerically, we can demonstrate the best of the three controls by using these choices of the feedback with the Bayesian control with conditional data, discussed in details down below. 

\section{Effect of input state}
\begin{figure*}[t]
\centering
    \includegraphics[width=0.9\textwidth]{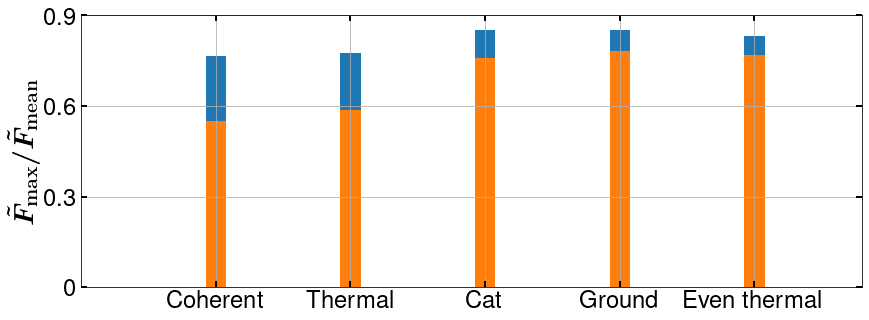}
    \caption{Effect of different starting conditions for solving the stochastic master equation on learning of the agent for the optimal value of measurement rate at $\Gamma=0.1$ with measurement records as control, the maximum (blue) and mean (orange) of 10 successive deterministic episodes of the trained agents are shown.}
    \label{fig:effect_of_inputs}
\end{figure*}

It is found that the choice of initial state for the time evolution via the stochastic master equation has a profound effect on how well the DRL agent can achieve its goal to bring the system to the ground state. The  maximum (blue) and mean (orange) fidelity of the trained agents for different initial states are shown in Fig.~\ref{fig:effect_of_inputs}. When the initial state, $\rho(0)$ is a thermal or coherent state, the DRL converges to an average fidelity of about 60\%. However, if we use a small cat state or the ground state of the DW itself, the agent is able to achieve a mean trained fidelity of over 80\% with noisy measurement data.  We achieve similar high performance if we start with a thermal state projected on to even parity.

\section{Effect of measurement efficiency and decoherence}

\begin{figure*}[h]
    \centering
    \includegraphics[width=0.9\textwidth]{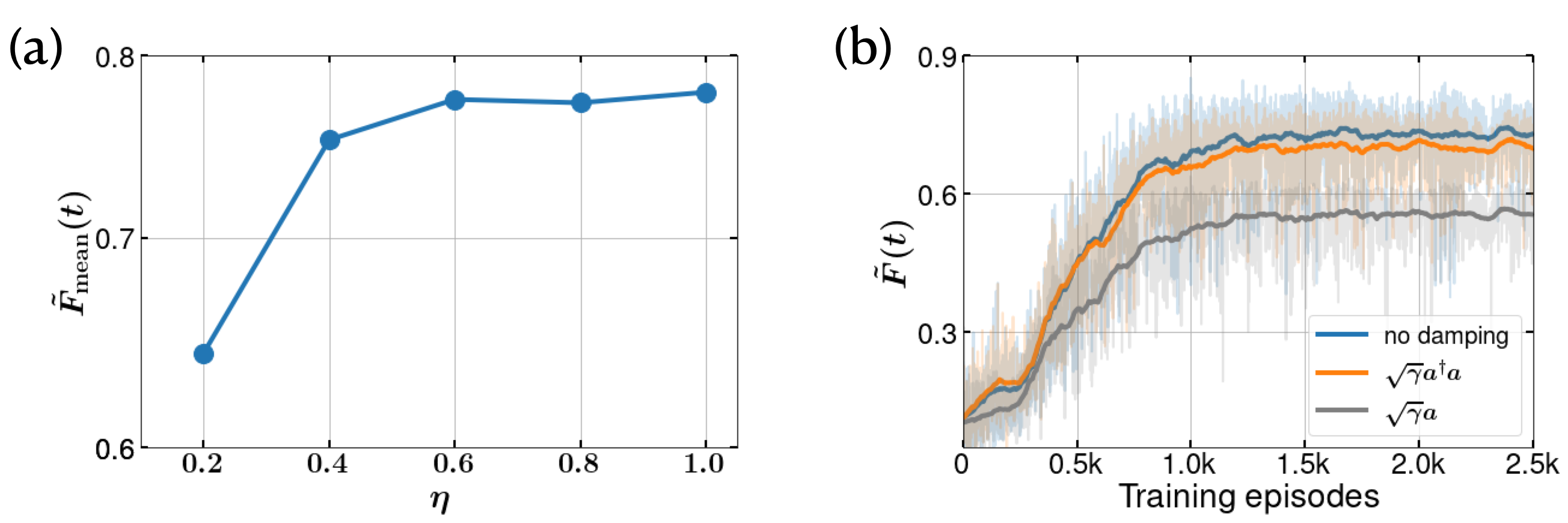}
    \caption{(a) Performance of the DRL is shown as the net mean fidelity  achieved by a trained RL-agent as a function of measurement efficiency. (b) The variation of net fidelity (over each episode) with (orange:  damping operator = $\sqrt{\gamma} a^\dagger a$ and gray:  damping operator = $\sqrt{\gamma} a^\dagger$), and without (blue) the presence of environment damping in the training process. The damping rate chosen for these simulations is  $\gamma =0.1$. }
    \label{fig:effect_of_damping_and_efficiency}
\end{figure*}

We benchmark the mean fidelity achieved as a function of the measurement efficiency, $\eta$ in the Fig.~\ref{fig:effect_of_damping_and_efficiency}(a). We find that the DRL reveals robust control for $\eta \geq 50\%$. In Fig.~\ref{fig:effect_of_damping_and_efficiency}(b), the learning of the DRL agent is compared with the one without environment damping (blue),in the presence of different collapse operators -- (i) $\sqrt \gamma a$, and (ii) $\sqrt \gamma a^\dagger a$. Further improvement of the results could be achieved by implementing the suggestions mentioned at the end of the main manuscript.

\section{Generalizability of the trained model}
\begin{figure*}[h]
    \centering
    \includegraphics[width=0.5\textwidth]{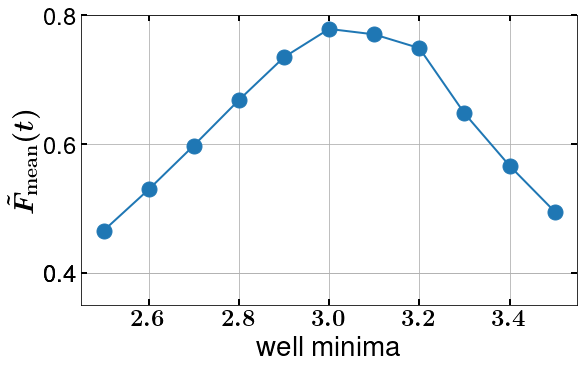}
    \caption{Testing the generalizability of the trained agent for a changes in the double-well model parameter (here the positions of the double-well minima). The model is trained originally for minima position at $x=\pm 3$. The loss of generality is due to the fact that the agent is trained explicitly based on the measurement records which changes as we change the parameters of the double-well. This suggests that the agent needs to be explicitly re-trained for a significant change in the parameters of the double-well.  }
    \label{fig:generalizability}
\end{figure*}
Since the DRL is trained explicitly on the measurement data that varies with the double-well parameters (for example, the position of the well minima), measurement strength ($\Gamma$) and measurement time interval ($\delta t$), the generalizability of a trained model is poor across any significant changes of those parameters (see Fig.~\ref{fig:generalizability}). However, this is not the case for the initial state $\rho(0)$. Although the choice of the type of $\rho(0)$ has a significant effect on the mean fidelity achieved by the agent, we found that a trained model with a given $\rho(0)$ (say coherent state) yields exactly the same behavior for others (say even-thermal state) and vice versa (see Fig.~\ref{fig:effect_of_inputs}), which means that choosing a better prediction (or a similar basis with respect to the target state) would lead to better net fidelity, without the need to retrain the model for different initial states. 

\section{Fidelity as the reward function}
\begin{figure*}[h]
    \centering
    \includegraphics[width=0.9\textwidth]{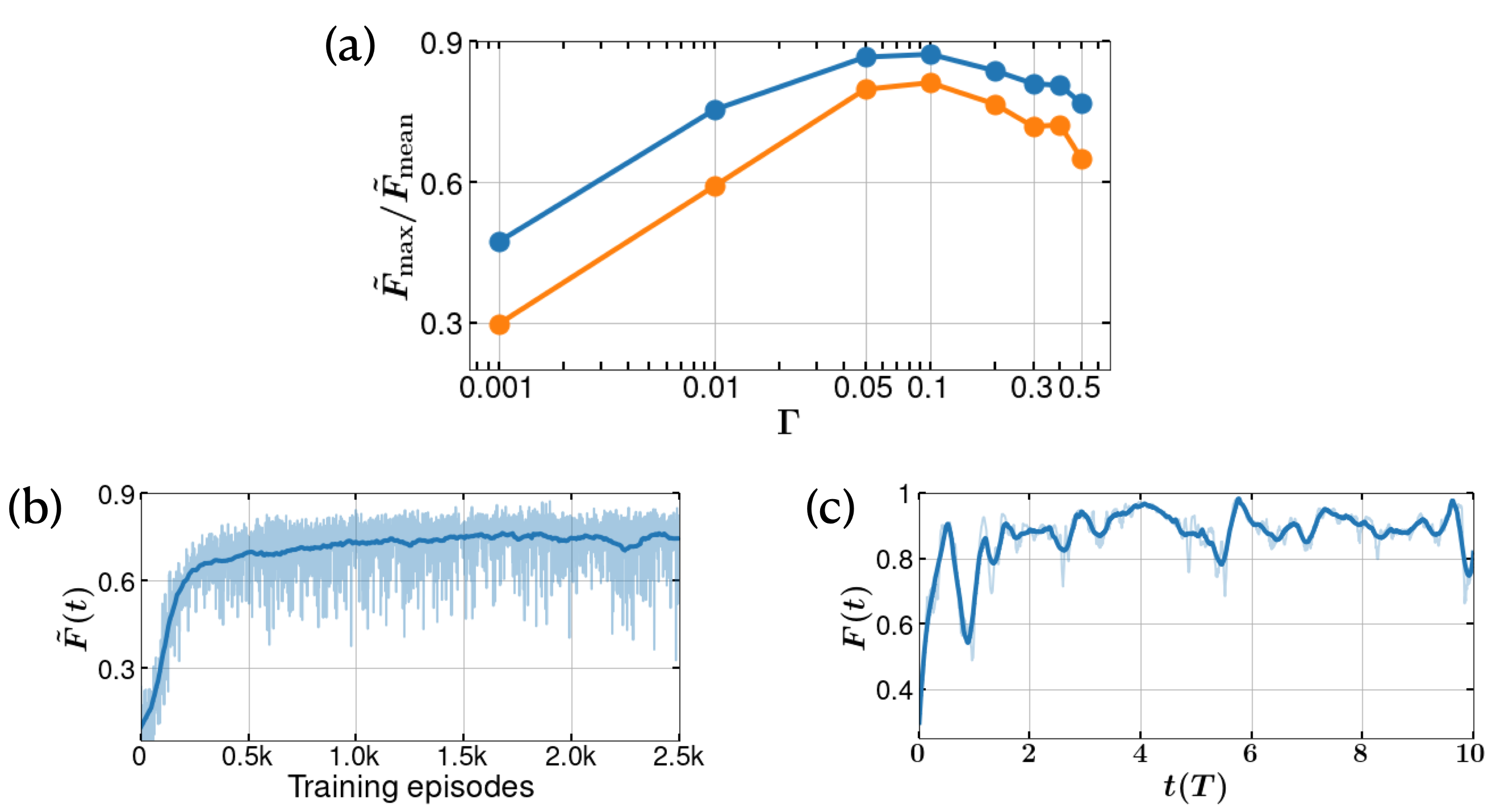}
    \caption{Fidelity as the reward function: (a) Effect of measurement rates, $\Gamma$, on learning of the agent, the maximum (in blue) and mean (in brown) fidelity out of 100 successive deterministic episodes of the trained agents are shown, (b) The learning of the agent for $\Gamma=0.1$ and (c) the variation of fidelity in a given episode of the trained agent. }
    \label{fig:supp_fidelity_as_control}
\end{figure*}

It is possible to choose fidelity as the reward function instead of measurement current discussed in the main text of the paper, shown in Fig.~\ref{fig:supp_fidelity_as_control}. However, fidelity is not directly accessible in real time in a laboratory experiment, and the following analysis is included for the sake of clarity and helpful comparison. When the DRL agent uses fidelity as a reward function we see that the learning behaviour, shown in  (Fig.~\ref{fig:supp_fidelity_as_control}(a)), as a function of the measurement rate $\Gamma$, essentially mirrors the behaviour seen when the measurement current is used instead as a reward function (shown in Fig. 2(a) in the main text of the paper). In Fig.~\ref{fig:supp_fidelity_as_control}(b) the episodic mean reward $\tilde F(t)$ evolution in time during the training of the agent when $\Gamma=0.1$ [light(dark) blue includes(averages) noise]. In Fig.~\ref{fig:supp_fidelity_as_control}(c) the fidelity variation of a given random episode of the training agent is shown.

\section{Bayesian  vs. DRL control}
Here we benchmark our result against the state-based Bayesian feedback protocol (where the feedback is based on an estimate of the state) as proposed by Doherty et al.,~\cite{Doherty1999Oct, Stockton2004Aug}. In the context of the present work, the protocol is simplified to provide feedback of the form $\mathcal{F}(t)= -(\langle {x}_c^2\rangle(t) - 3^2) \times (xp + px)$ where $\langle x_c^2 \rangle (t)$ denotes the conditional mean of the observable $x^2$.  However, $\langle x_c^2 \rangle (t)$ is not a quantity directly accessible in real experiments, and the above feedback condition should be replaced by the measurement current, $I(t)$. We find that when Bayesian feedback is driven by $I(t)$, it shows very little control over the dynamics. We performed a numerical simulation with 1000 copies of the system (ensemble), all evolving under feedback based on the mean of the measurement currents during each time step. The performances of these two approaches is compared in  Table \ref{table:bayesian_vs_drl}

\begin{table}[h]
\centering
\caption{Comparison of the Bayesian and DRL control towards finding the ground state of the double well potential.  We list a comparison of the mean fidelities achieved when the controllers are driven  either by the conditional mean of  $\langle x_c^2\rangle(t)$  [which are not readily available during a live experiment], or by the measurement current $I(t)$, which includes measurement noise. In addition we include cases with decoherence of two forms, either amplitude damping decoherence at a rate $\gamma=0.1$ or dephasing decoherence at the same rate.  The Bayesian control for the measurement current is done over an ensemble of 1000 trajectories. We color blue the higher fidelity result for clarity.}
\vspace{.5cm}
\begin{tabular}{|c|c|c|}

\hline
Feedback basis                      & Bayesian control (\% fidelity)  & DRL control (\% fidelity) \\
\hline
conditional means, $\langle x_c^2\rangle$(t)                                         & \cellcolor{blue!25}   85\%       &   83\%  \\
 $\langle x_c^2\rangle$(t) with damping $\sqrt \gamma a$           &    62\%     &  \cellcolor{blue!25}   63\% \\
 $\langle x_c^2\rangle$(t) with damping $\sqrt \gamma a^\dagger a$ &    76\%       & \cellcolor{blue!25}  79\%  \\
  & & \\
measurement current, $I(t)$         &  42\%       & \cellcolor{blue!25}  77\%  \\
$I(t)$ with damping $\sqrt \gamma a$                              &    40\%      & \cellcolor{blue!25} 58\%   \\
$I(t)$ with damping $\sqrt \gamma a^\dagger a$                    &    41\%     &\cellcolor{blue!25}   72\%  \\
\hline
\end{tabular}
\label{table:bayesian_vs_drl}
\end{table}

We see from \ref{table:bayesian_vs_drl} that in order that the Bayesian control performs well only when it is provided the control with conditional mean data. When used with measurement currents, it, however, shows no control at all. Even when it is used with an ensemble of 1000 identical systems, it does not find any control beyond 50\%. We found from our analysis that it is very important to provide the Bayesian controller with very accurate estimation of the mean currents that does not deviate  significantly from the conditional means. The presence of a bad data point often throws the system into a non-recoverable state resulting in much lower fidelities. DRL, on the other hand, has a brain, and hence can recover from such bad data points. 

It is also seen that when the Bayesian as well as the DRL controller is trained with conditional mean data, both yield comparable performances. It is interesting to observe the significant drop in achievable fidelity in presence of the environment damping $\sqrt{\gamma} a$, but not in presence of $\sqrt{\gamma} a^\dagger a$, environment dephasing. Thus, the drop of the performance of the DRL agent in presence of an environment damping $\sqrt{\gamma} a$ can be attributed to the reduction of the indistinguishably of the target  state of the DW potential in the presence of  $\sqrt{\gamma} a$ decoherence.

Finally, we found that the choice of the $xp+px$ feedback over $x^2$ and $p^2-x^2$, discussed above, can be demonstrated using Bayesian feedback with conditional mean data. We found a feedback of the form $x^2$ could never control the dynamics, while with $p^2-x^2$ achieves a mean fidelity of $\sim 60\%$. The mean fidelity of 85\% shown shown in table~\ref{table:bayesian_vs_drl} could only be obtained when used with the feedback $xp + px$.


%

\end{document}